\documentclass{aa}
\bibpunct{(}{)}{;}{a}{}{,} 
\usepackage{graphicx}
\usepackage[varg]{txfonts}
\usepackage{color}
\usepackage{footmisc}
\usepackage[utf8]{inputenc}
\usepackage[T1]{fontenc}    
\usepackage{hyperref}       
\begin{document}
\title{Polarization properties of changing-look active galactic nuclei: NGC~1365 and NGC~2992\thanks{Based on observations made with the William Herschel telescope operated on the island of La Palma by the Isaac Newton Group of Telescopes in the Spanish Observatorio del Roque de los Muchachos of the Instituto de Astrof{\'\i}sica de Canarias, and data obtained from the European Southern Observatory Science Archive Facility under programme ID~0101.B-0530.}}
\author{D. Hutsem\'ekers\inst{1},
        F. Marin\inst{2},
        B. Ag{\'\i}s Gonz\'alez\inst{3},
        J.-A. Acosta Pulido\inst{4,5},
        M. Kokubo\inst{6}
        }
\institute{
    Institut d'Astrophysique et de G\'eophysique, Universit\'e de Li\`ege, All\'ee du 6 Ao\^ut 19c, 4000 Li\`ege, Belgium
    \and
    Universit\'e de Strasbourg, CNRS, Observatoire Astronomique de Strasbourg, UMR 7550, F-67000 Strasbourg, France
    \and
    Institute of Astrophysics, Foundation for Research and Technology-Hellas (FORTH), Heraklion 70013, Greece
    \and
    Instituto de Astrofisica de Canarias, Calle Via Lactea, s/n, E-38205 La Laguna, Tenerife, Spain
    \and
    Departamento de Astrofisica, Universidad de La Laguna, E-38205 La Laguna, Tenerife, Spain
    \and
    National Astronomical Observatory of Japan, 2-21-1 Osawa, Mitaka, Tokyo 181-8588, Japan}
\date{Received ; accepted: }
\titlerunning{Polarization of the changing-look AGNs in NGC 1365 and NGC 2992} 
\authorrunning{D. Hutsem\'ekers et al.}
\abstract{
Changing-look active galactic nuclei (CLAGNs) represent a rare class of AGNs that undergo transitions from type~1 (characterized by the presence of broad emission lines in their spectra) to type~2 (absence of broad emission lines) or vice versa, over timescales ranging from months to years. Such brief periods of variation can significantly constrain the physical processes involved in the accretion of matter onto the AGN central supermassive black hole. Since normal type~1 and type~2 AGNs are known to show different polarization properties, detailed investigations of the CLAGN polarization can shed light on the underlying mechanisms responsible for the changing-look phenomenon.
In this paper we present new (spectro)polarimetric observations of two changing-look AGNs located in the core of the inclined spiral galaxies NGC~1365 and NGC~2992. Both AGNs are radio emitters, thereby enabling a comparison of their polarization to the radio jet axis, which defines the accretion disk geometry.
In the case of NGC~1365, the AGN shows polarization characteristics consistent with those observed in type~1 Seyferts, in particular polarization parallel to the radio jet. This intrinsic polarization is modified by the wavelength-dependent dichroic extinction that occurs in the galaxy bar and that rotates the polarization angle at the shortest wavelengths. Measuring the polarization of NGC~1365 when the AGN is in a type~2 state would be particularly interesting to see if its polarization becomes perpendicular to the radio axis, as observed in normal type~2 AGNs, or if it remains parallel.
NGC~2992, on the other hand, is so inclined that dichroic dust extinction in the disk completely dominates the polarization of the AGN, thus overwhelming  any polarization due to scattering. Consequently, the polarization properties remain essentially constant between the different AGN states, and the faint broad lines observed in the polarized flux are most likely not scattered light. Differential dilution between the continuum and the narrow-line polarizations can explain the unusually high polarization measured in the emission lines. 
}
\maketitle
%
%
%

\section{Introduction}
\label{sec:intro}

The changing-look phenomenon, which occurs in some active galactic nuclei (AGNs) and involves the disappearance or appearance of broad emission lines (BELs), is most often explained by a change in the accretion rate onto the supermassive black hole \citep{2015Lamassa,2017Sheng,2017Hutsemekers,2023Ricci,2023Temple,2025Duffy}. This change leads to a dimming or brightening of the light source. Thus, changing-look AGNs (CLAGNs) can be used to study  accretion physics. However, it is unclear whether the broad-line region (BLR), which is the origin of the BELs, simply stops shining in faint states or if it completely disappears and reappears. If the latter is true, we need to determine whether the equatorial scattering region, which is located close to the BLR and is responsible for the polarization of type 1 AGNs \citep{1994Goodrich,2000Young,2002Smith,2007Goosmann}, also disappears in faint states.

Imaging polarimetry revealed that the polarization degree of CLAGNs remains essentially unchanged when type~1 AGNs (those with BELs) transition to type 2 AGNs (those without BELs). This finding rejects the idea that the changing-look phenomenon is predominantly due to variable dust obscuration in the torus \citep{2017Hutsemekers,2019Hutsemekers}. Nevertheless, modeling the polarization behavior during the changes of look revealed interesting phenomena such as the presence of polarization echoes, or the possible existence of type~2 AGNs with polarization parallel to the radio axis \citep{2017Marin,2020Marin}, knowing that type~2 AGNs almost exclusively show polarization perpendicular to the radio axis \citep{1983Antonucci,1984Antonucci}. Therefore, spectropolarimetry of CLAGNs is an important tool for further understanding the changing-look phenomenon, ideally by gathering observations when the objects are in different states. In this paper, we present and discuss new polarization measurements of two CLAGNs located in the spiral galaxies NGC~1365 and NGC~2992, both of which show resolved radio emission that will allow us to compare the polarization with the radio axis. Hereafter, NGC~1365N and NGC~2992N denote the AGNs in the galaxies NGC 1365 and NGC 2992, respectively.

NGC~1365 is a supergiant barred spiral galaxy seen with an inclination $i \simeq 40 \degr$ \citep[see][for a detailed overview of NGC~1365]{1999Lindblad}. It has a redshift of $z = 0.0054$ and an angular-size distance of 22.6 Mpc (NED\footnote{NASA/IPAC Extragalactic Database, \\ https://ned.ipac.caltech.edu/}).  NGC~1365N regularly transitions between type~1 and type~2, with intermediate states, on a timescale of a few years \citep[e.g.,][]{2023Temple}.
NGC~2992 is a highly inclined ($i \simeq 70 \degr$) Sa galaxy \citep[see][for a summary of NGC~2992's main characteristics]{1999Felton}. It has a redshift $z = 0.0077$ and an angular-size distance of 38.7 Mpc (NED). NGC~2992N shows more modest changes of look, essentially between type~1.8 and type~2 \citep[e.g.,][]{2021Guolo}.

In Sect.~\ref{sec:obs} we describe the observations and their reductions. We then analyze the data and discuss the results for NGC~1365 and NGC~2992 in Sects.~\ref{sec:results1} and~\ref{sec:results2}, respectively. Our conclusions are presented in the last section.

\section{Observations and data reductions}
\label{sec:obs}

\subsection{Spectropolarimetry with  VLT + FORS2}

\begin{figure}[]
\centering
\resizebox{0.9\hsize}{!}{\includegraphics*{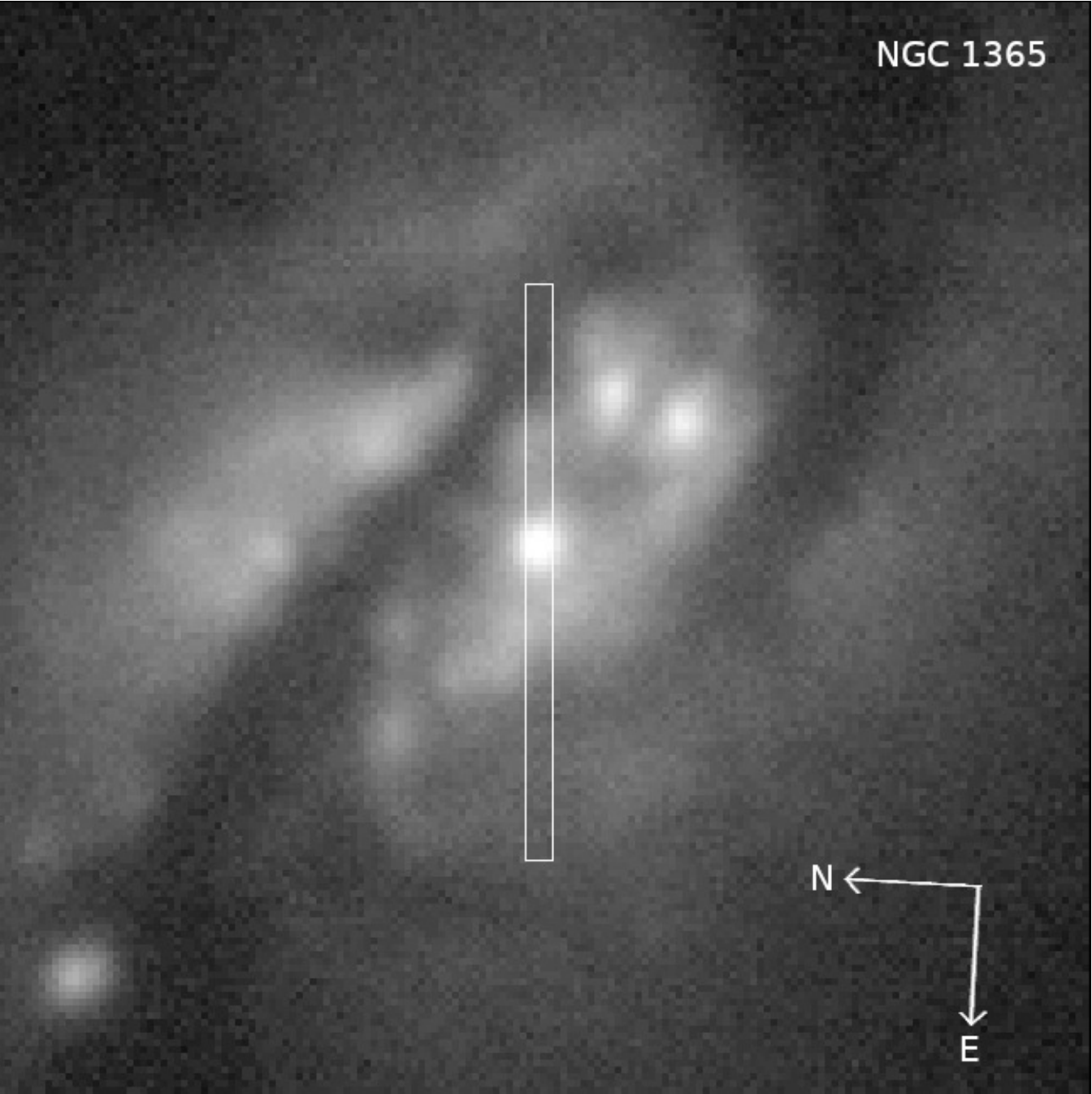}}
\caption{Visible image of the central part of NGC~1365 obtained during the FORS2 observations. The field size is 40\arcsec $\times$ 40\arcsec. The directions north and east are indicated. The position of the slit, centered on the AGN, is also shown.}
\label{fig:ngc1365}
\end{figure}

\begin{figure}[]
\centering
\resizebox{0.9\hsize}{!}{\includegraphics*{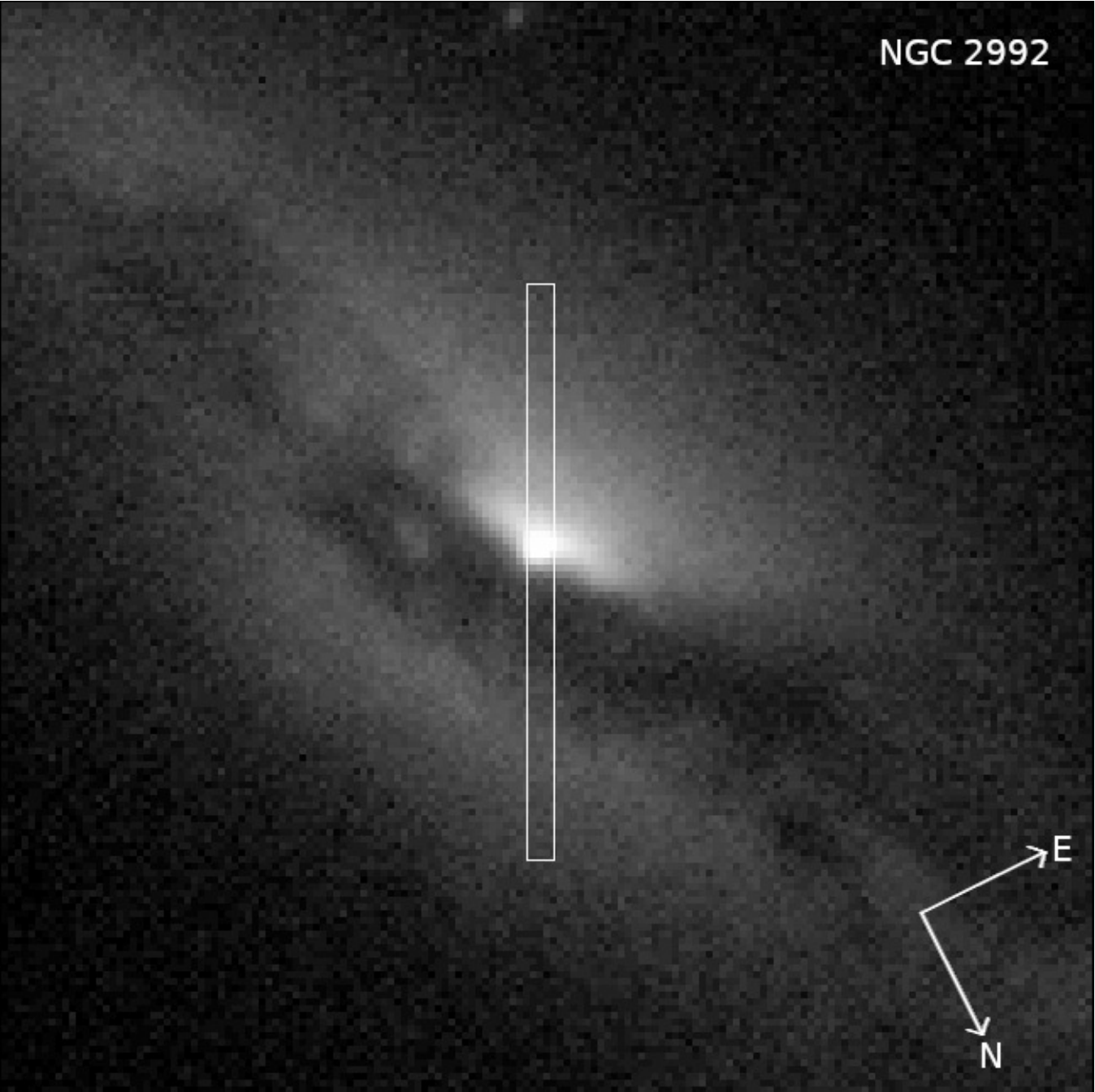}}
\caption{Same as Fig.~\ref{fig:ngc1365}, but for NGC~2992.}
\label{fig:ngc2992}
\end{figure}

Spectropolarimetric observations of NGC~2992N and NGC~1365N were retrieved from the European Southern Observatory (ESO) archive (program ID: 0101.B-0530; PI: M.~Kokubo\footnote{Spectropolarimetry of a large sample of type 1.5, 1.8, and 1.9 Seyfert galaxies will be discussed elsewhere (Kokubo et al., in preparation).}). They were obtained on April 20 and July 24, 2018, respectively, using the Very Large Telescope (VLT) equipped with the focal reducer and low dispersion spectrograph FORS2, which is mounted at the Cassegrain focus of Unit Telescope \#1 (Antu). Linear spectropolarimetry was performed by inserting in the beam a Wollaston prism that splits the incoming light rays into two orthogonally polarized beams separated by 22$\arcsec$ on the CCD detector.\footnote{\label{note1}FORS2 User Manual, VLT-MAN-ESO-13100-1543,\\ https://www.eso.org/sci/facilities/paranal/instruments.html}  The grism 300V was used with the order sorting filter GG435 to cover the spectral range 4500$-$8700~\AA. The average resolving power was R $\simeq$ 440 for a 1\arcsec\ slit. Four frames were obtained with the half-wave plate (HWP) rotated at four different position angles, 0\degr, 22.5\degr, 45\degr, and 67.5\degr. The MOS slit was 1\arcsec~wide and 20\arcsec~long. It was centered on the nucleus and oriented along the parallactic angle (Figs.~\ref{fig:ngc1365} and~\ref{fig:ngc2992}). The spatial scale on the detector was 0$\farcs$25 per pixel with a 2$\times$2 binning. The seeing was around 0.8\arcsec\ during the observations of NGC~2992N and 1.3\arcsec\ during the observations of NGC~1365N. 

Raw frames were first processed to remove cosmic ray hits using the Python implementation of the lacosmic package \citep{2001VanDokkum,2012VanDokkum}. The ESO FORS2 pipeline\footnote{ESO Pipeline User Manual, VLT-MAN-ESO-19500-4106,\\ https://www.eso.org/sci/software.html} was then used to get images with two-dimensional spectra rectified and calibrated in wavelength. The one-dimensional spectra were extracted using a 3$\farcs$25 (13 pixel)-long subslit centered on the nucleus. This subslit was as small as possible to minimize contamination from the host galaxy, taking into account the seeing. Small changes of the aperture do not affect the results. The sky spectrum was estimated from adjacent MOS strips and subtracted from the nucleus spectrum.  The normalized Stokes parameters $q(\lambda)$ and $u(\lambda)$, the linear polarization degree $p(\lambda)$, the polarization position angle $\theta(\lambda)$, and the total flux density $F(\lambda)$ were then computed from the ordinary and extraordinary spectra according to standard recipes (FORS2 Manual\footref{note1}; \citealt{2025Marin}, \citealt{2009Bagnulo}).  The direct spectrum $F(\lambda)$ was corrected for extinction and calibrated in flux using a master response curve.  The spectra were corrected for the rotator angle and for the retarder plate zero angle provided in the FORS2 manual so that $\theta(\lambda)$ = 0$\degr$ (90$\degr$) corresponds to the north (east) direction.  The uncertainties were estimated by propagating the photon and readout noise. The polarization degree $p_d(\lambda)$ was debiased using the modified asymptotic estimator of \citet{2014Plaski}. Polarized (BD$-12\degr 5133$) and unpolarized (WD1620$-$391) standard stars \citep{2007Fossati} were observed and reduced as the targets to check the whole reduction process. The instrumental polarization was found smaller than 0.06\%. The final polarization spectra, binned over the spectral resolution element (4 pixels $\simeq$ 3.3 \AA), are shown in Figs.~\ref{fig:spola_ngc1365} and~\ref{fig:spola_ngc2992}.

\subsection{Imaging polarimetry with  WHT + ISIS}

Linear polarization data were obtained for NGC~2992N on January 2, 2020, using the Intermediate dispersion Spectrograph and Imaging System (ISIS) mounted at the Cassegrain focus of the 4.2m William Herschel Telescope (WHT) located at the Roque de los Muchachos Observatory (program ID: 170-WHT14/19B; PI: J.A.~Acosta Pulido). Observations were done through the blue arm and the Sloan Gunn $g$ filter: ING filter \#218; $\lambda_c$ = 4844 \AA , full width at half-maximum (FWHM) = 1280 \AA , with ISIS in its imaging polarimetry mode.\footnote{ISIS User's Manual,\\ https://www.ing.iac.es/astronomy/observing/manuals} Polarimetry with ISIS was performed by using a calcite Savart plate that  produces two orthogonally polarized images in 6.4\arcsec  wide strips separated by 7.7\arcsec. Four exposures with the HWP rotated at 0\degr, 22.5\degr, 45\degr, and 67.5\degr\ were secured. The spatial scale on the detector was 0$\farcs$2 per pixel with a 1$\times$1 binning. The sky-subtracted integrated intensities of the orthogonally polarized images of the object were measured using a circular aperture of diameter 3.2\arcsec\ (16 pixels) to minimize the host contamination, although slightly larger apertures did not significantly change the polarization measurements. The normalized Stokes parameters, the polarization degree, and the polarization position angle  were computed using the recipes described in \citet{2018Hutsemekers}.  Polarized (BD$+25\degr 727$ = HD283812) and  unpolarized (G191B2B, and GD319) standard stars \citep{1990Turnshek,2016Slowikowska} were observed to correct for the chromatic dependence of the HWP zero-angle and estimate the instrumental polarization ($p \leq 0.1 \%$). The final polarization of NGC~2992N on January 2, 2020, was found to be $p = 3.72 \pm 0.03 \%$ with $\theta = 36 \pm 1 \degr$.

\subsection{Interstellar polarization}

To estimate the possible contamimation due to interstellar polarization in our Galaxy, we searched for stars as close as possible to our targets in the catalog of optical starlight polarization compiled by \citet{2025Panopoulou}. For NGC~1365N, the two closest stars are at angular distances of 2.87 and 3.44 degrees. Their white-light polarizations are $p$ = 0.023$\pm$0.016\% at $\theta$ = 104$\pm$26\degr\ and $p$ = 0.013$\pm$0.002\% at $\theta$ = 6$\pm$5\degr. For NGC~2992N, the two closest stars are at angular distances of 1.31 and 3.25 degrees. Their white-light polarizations are $p$ = 0.061$\pm$0.014\%  at $\theta$ = 25$\pm$7\degr\ and $p$ = 0.057$\pm$0.022\%  at $\theta$ = 74$\pm$12\degr. In both cases, the starlight polarizations are very small, and significantly smaller than the polarization of our targets ($\geq 1\%$).  We thus considered the contamination by interstellar polarization in our Galaxy to be negligible for both AGNs.

\begin{figure}[]
\centering
\resizebox{\hsize}{!}{\includegraphics*{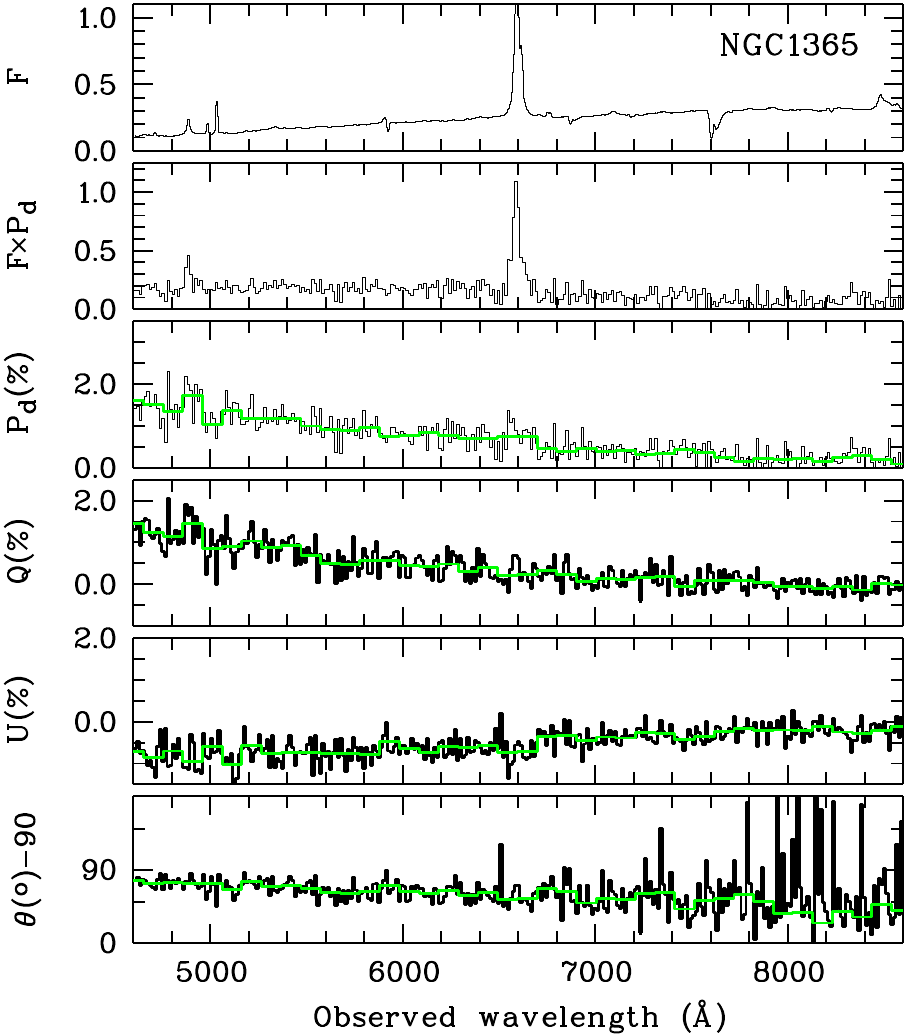}}
\caption{Spectropolarimetry of NGC~1365N. From top to bottom: Direct and  polarized fluxes in arbitrary units,  debiased polarization degree in percent,  normalized Stokes parameters $q$ and $u$ in percent, and  polarization position angle in degrees (from which 90\degr\ were subtracted for clarity of the plot). The spectra in green were binned over  30 spectral pixels (100 \AA ) to increase the signal-to-noise ratio.}
\label{fig:spola_ngc1365}
\end{figure}

\section{Analysis and results: NGC~1365}
\label{sec:results1}

\subsection{Direct spectrum}

\begin{figure}[t]
\centering
\resizebox{0.9\hsize}{!}{\includegraphics*{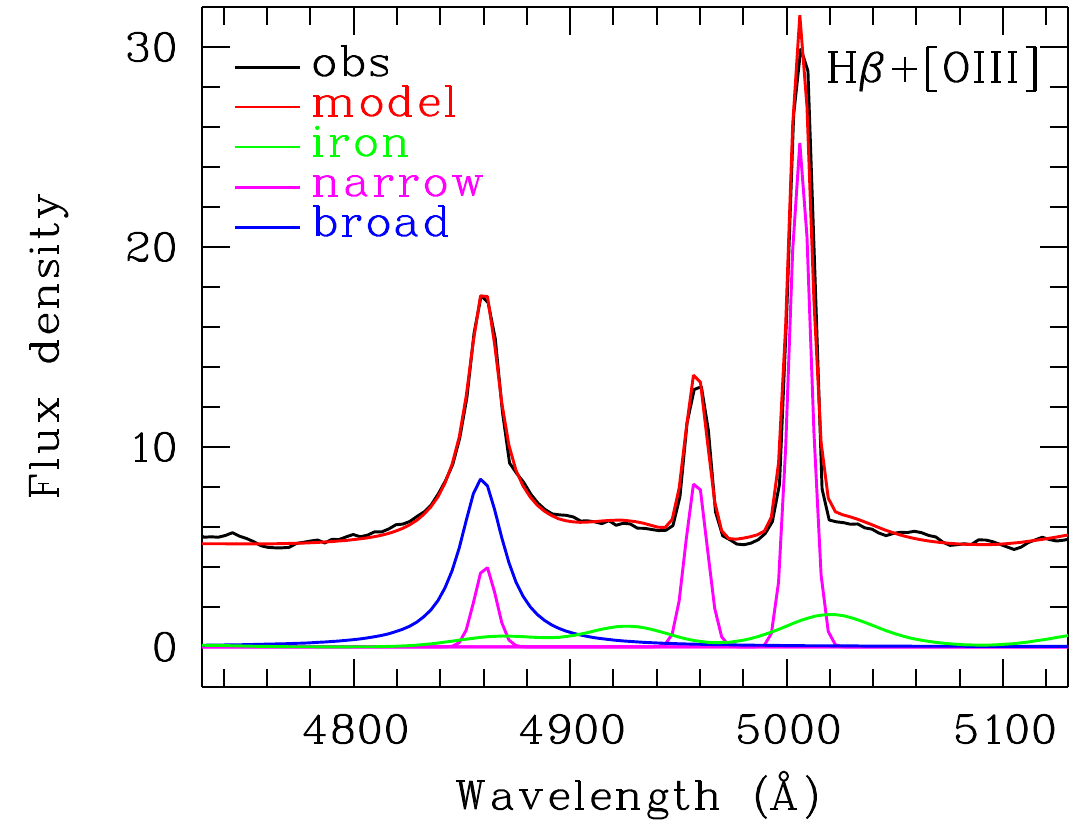}}
\caption{Fit of narrow- and broad-line profiles to the H$\beta$ and [\ion{O}{iii}] emission lines observed in the direct spectrum of NGC~1365N. The continuum and the \ion{Fe}{ii} emission (in green) were previously subtracted. The observed and the modeled spectra are superimposed and shifted along the y-axis for clarity. The flux density is given in arbitrary units.}
\label{fig:ngc1365_hbeta}
\end{figure}

\begin{figure}[t]
\centering
\resizebox{0.9\hsize}{!}{\includegraphics*{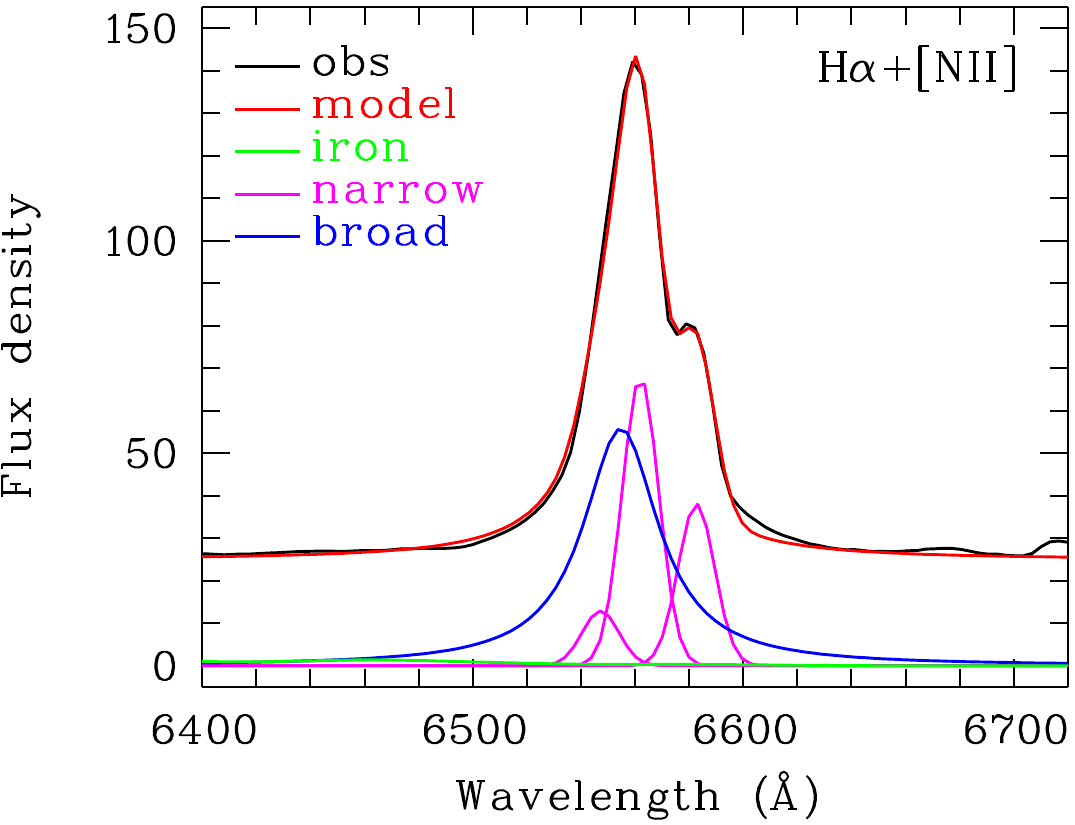}}
\caption{Same as Fig.~\ref{fig:ngc1365_hbeta}, but for the H$\alpha$ and [\ion{N}{ii}] emission lines.}
\label{fig:ngc1365_halpha}
\end{figure}

The direct spectrum of NGC~1365N shows a red continuum with prominent Balmer (H$\beta$ and H$\alpha$), [\ion{O}{iii}] $\lambda\lambda$~4959,5007, and  [\ion{N}{ii}] $\lambda\lambda$~6548,6584 emission lines. To separate the broad and narrow lines, we first subtracted the continuum from the direct spectrum before fitting a \ion{Fe}{ii} emission model using FANTASY\footnote{Fully Automated pythoN tool for AGN Spectra analYsis,\\ https://fantasy-agn.readthedocs.io/en/latest/} \citep{2020Ilic,2022Rakic,2023Ilic}. After subtraction of the \ion{Fe}{ii} model, we fitted the narrow lines with Gaussian profiles, all with same velocity width, and the broad H$\alpha$ and H$\beta$ emission lines with Lorentzian profiles assumed to have the same width in velocity. The results are shown in Figs.~\ref{fig:ngc1365_hbeta} and~\ref{fig:ngc1365_halpha}. The FWHM of the narrow lines (unresolved)  is around 700 km~s$^{-1}$, while the FWHM of the broad Balmer lines is around 1500 km~s$^{-1}$, in agreement with previous studies \citep{1980Veron,1982Edmunds}. The broad H$\alpha$ appears blueshifted by $\sim$ 350 km~s$^{-1}$ with respect to the narrow lines, possibly indicating outflowing material. On July 24, 2018, NGC~1365N was thus clearly in a type~1 state, more precisely type~1.5 since $ 1 < F_{[\ion{O}{iii}]5007} / F_{H_{\beta}} \leq 4 $ according to the classification of \citet{1992Whittle}. The observed H$\alpha$/H$\beta$ broad emission line ratio is equal to 9. Adopting a theoretical ratio of 3.06 \citep{2008Dong} and following the procedure described in \citet{2013Momcheva}, we derived $E(B-V)$ = 0.9 and $A_V$ = 3.6 mag. 

\citet{2023Temple} summarized the rich changing-look history of NGC~1365N and showed relatively recent spectra from the BAT AGN spectroscopic survey. In 2010, NGC~1365N was in type~2 with only narrow lines. In 2013 and 2017, NGC~1365N was in a type~1 state, with broad Balmer lines clearly observed. In 2021, NGC~1365N was back in type~2. Our results indicate that NGC~1365N was still in a type~1 state in 2018, before transiting to the type~2 state observed in 2021. In 2024, NGC~1365N was again in a type~1 state (Appendix~\ref{appa}).

\subsection{Polarization}

The polarization of NGC~1365N (Fig.~\ref{fig:spola_ngc1365}) smoothly changes with the wavelength. We measured $p = 1.43 \pm 0.08 \%$ with $\theta = 162 \pm 2 \degr$ when integrating $q$ and $u$ over the 4620--4820~\AA\ wavelength range, and $p = 0.30 \pm 0.04 \%$ with $\theta = 129 \pm 5 \degr$ over the 8350--8550~\AA\ wavelength range; these wavelength windows were arbitrarily chosen at the blue and red ends of the spectrum. \citet{1990Brindle} reported B, V, H broadband polarimetry of NGC~1365N obtained on October 6, 1983: $p_B = 1.19 \pm 1.16 \%$ with $\theta_B = 164 \pm 28 \degr$, $p_V = 0.91 \pm 0.18 \%$ with $\theta_V = 157 \pm 6 \degr$, and $p_H = 0.11 \pm 0.16 \%$ with $\theta_H = 120 \pm 42 \degr$. Although the uncertainties are large, the trends with the wavelength are similar. NGC~1365N was reported to be in a type~1 state in 1979 and 1988 \citep{1982Edmunds,1994Schulz}, and thus most probably also in a type~1 state in 1983.

The broad Balmer emission lines are present in the polarized flux, with widths comparable to the widths of the broad lines measured in the direct flux (Figs.~\ref{fig:ngc1365_hbeta} and \ref{fig:ngc1365_halpha}). The [\ion{O}{iii}] narrow lines have a lower polarization degree and are not detected in the polarized flux. The polarization integrated over the H$\alpha$ line is $p = 0.89 \pm 0.05 \%$ with $\theta = 148 \pm 2 \degr$. The average polarization of the continuum from similar wavelength bins on both sides of H$\alpha$ is $p = 0.53 \pm 0.04 \%$ with $\theta = 153 \pm 2 \degr$. The polarization degree in H$\alpha$ is thus slightly higher than in the adjacent continuum, and the polarization angle marginally lower.

The rotation of the polarization angle with the wavelength and the fact that the broad-line polarization slightly differs from that of the adjacent continuum suggest that two polarization mechanisms are at work. \citet{1995Sandqvist} reported the presence of a radio jet that originates from the nucleus, at a position angle of 125$\degr$. This angle is remarkably similar to the polarization angle measured at the red end of our spectra, suggesting parallel polarization due to scattering, as observed in most type~1 Seyferts \citep[e.g.,][]{2002Smith}. The rotation of the polarization angle and the polarization degree increase toward bluer wavelengths could then be due to dichroic extinction in the bar of NGC~1365 through which the Seyfert nucleus is seen (Fig.~\ref{fig:ngc1365}). The reddening and extinction of the direct spectrum supports this interpretation. \citet{2005Beck} found that the magnetic field essentially follows the dark dust lanes roughly parallel to the bar (Fig.~\ref{fig:ngc1365}), except in the central region where the magnetic field deviates by about 70\degr\ to become roughly perpendicular to the bar. Since the bar has a position angle of about 92\degr, the position angle of the magnetic field in the central region superimposed on the nucleus is thus around 162\degr, which is precisely the polarization angle measured at the blue end of our spectra. Dichroic extinction by dust grains will thus generate an additional polarization with a polarization angle of 162\degr, and produce a rotation of the AGN polarization toward that angle as the wavelength decreases. To be more quantitative, the Stokes parameters add as follows:
\begin{eqnarray}
q_{\text{obs}} = q_{\text{sca}} + q_{\text{dic}},\\
u_{\text{obs}} = u_{\text{sca}} + u_{\text{dic}}.%
\end{eqnarray}
Here the subscripts obs, sca, and dic respectively refer to the observed polarization, the polarization generated by scattering with polarization angle $\theta_{\text{sca}}$, and the polarization generated by dichroism with polarization angle $\theta_{\text{dic}}$. These equations can be rewritten as
\begin{eqnarray}
\cos(2 \, \theta_{\text{obs}}) & = & r_{\text{sca}} \; \cos(2 \, \theta_{\text{sca}}) + r_{\text{dic}} \; \cos (2 \, \theta_{\text{dic}}),\\
\sin(2 \, \theta_{\text{obs}}) & = & r_{\text{sca}} \; \sin(2 \, \theta_{\text{sca}}) + r_{\text{dic}} \; \sin (2 \, \theta_{\text{dic}}),
\end{eqnarray}
where $r_{\text{sca}} = p_{\text{sca}} / p_{\text{obs}}$ and  $r_{\text{dic}} = p_{\text{dic}} / p_{\text{obs}}$. Given $\theta_{\text{obs}}$ = 162\degr\ at $\sim$4700~\AA\ and  $\theta_{\text{obs}}$ = 129\degr\ at $\sim$8500~\AA, these equations are trivially solved assuming $\theta_{\text{sca}}$ = 125\degr\ and  $\theta_{\text{dic}}$ = 162\degr, so that $r_{\text{dic}} \simeq $ 1 at $\sim$4700\AA\ and  $r_{\text{sca}} \simeq $ 1 at $\sim$8500\AA. More interesting is the fact that this is the only valid solution. Assuming perpendicular scattering ($\theta_{\text{sca}}$ = 35\degr) and/or  dust polarization parallel to the bar ($\theta_{\text{dic}}$ = 92\degr)
 gives unphysical solutions, i.e., negative values for  $r_{\text{sca}}$ or  $r_{\text{dic}}$.

In Appendix~\ref{appb}, we tested this interpretation further by fitting the observed polarization in the $q$, $u$ plane with a simple model in order to isolate the scattering component. As shown in Fig.~\ref{fig:ngc1365_uq}, the model can reproduce the observations, thus supporting the scattering plus dichroic extinction explanation. Furthermore, the polarization corrected for dichroic extinction remains constant across the whole wavelength range, suggesting that it originates from electron scattering.

In conclusion, the AGN in NGC~1365 shows the typical polarization of type~1 Seyferts with polarization parallel to the radio jet, modified by wavelength-dependent dichroic extinction in the galaxy bar. Dilution by the host unpolarized light can also contribute to the change of the polarization degree with wavelength, the host contamination being stronger in the red, but it does not affect the polarization angle. Finally, our results confirm that the magnetic field in the central region of the bar is deflected and roughly perpendicular to the bar.

\begin{figure}[]
\centering
\resizebox{\hsize}{!}{\includegraphics*{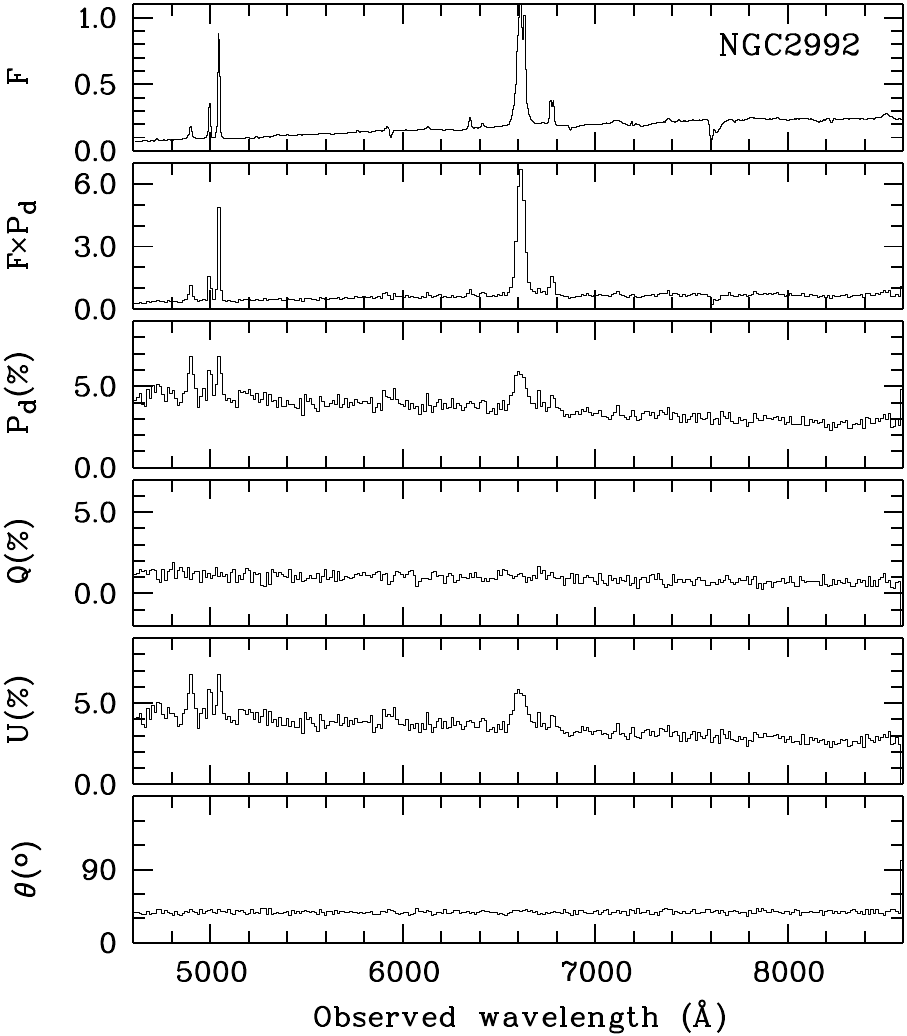}}
\caption{Spectropolarimetry of NGC~2992N. From top to bottom: Direct and the polarized fluxes in arbitrary units,  debiased polarization degree in percent,  normalized Stokes parameters $q$ and $u$ in percent, and  polarization position angle in degrees.}
\label{fig:spola_ngc2992}
\end{figure}

\begin{figure}[t]
\centering
\resizebox{0.9\hsize}{!}{\includegraphics*{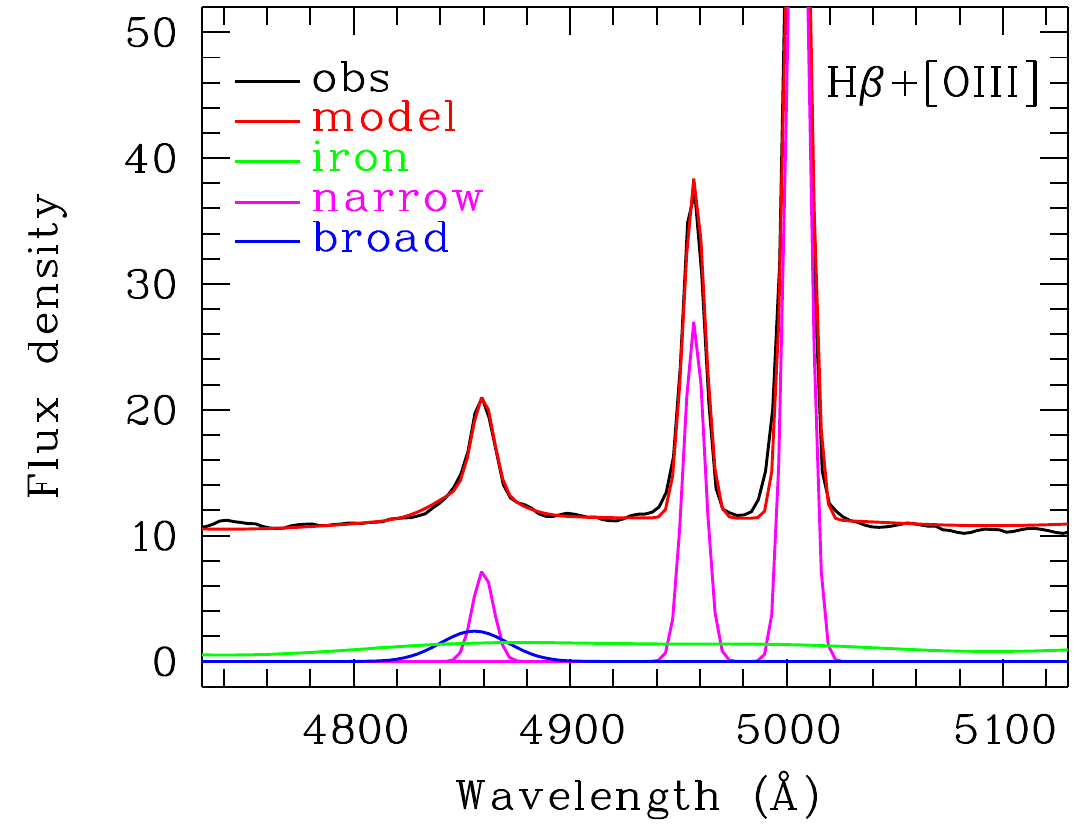}}
\caption{Same as Fig.~\ref{fig:ngc1365_hbeta}, but for NGC~2992N.}
\label{fig:ngc2992_hbeta}
\end{figure}

\begin{figure}[t]
\centering
\resizebox{0.9\hsize}{!}{\includegraphics*{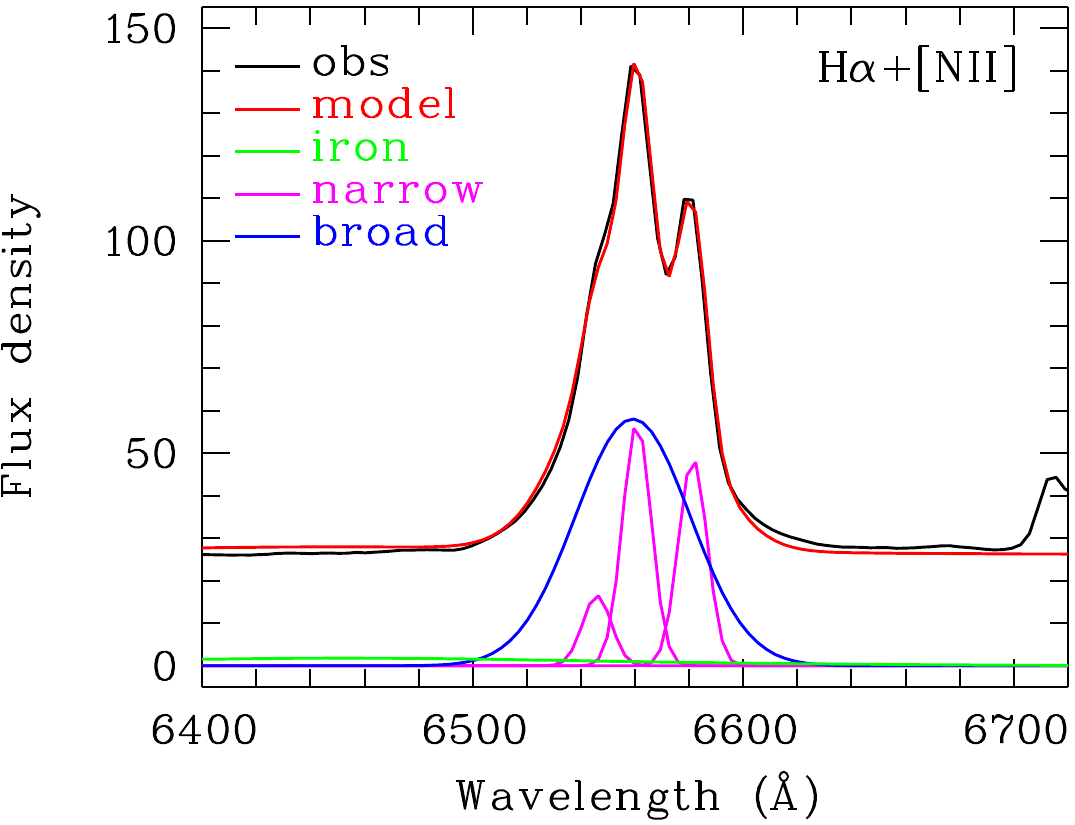}}
\caption{Same as Fig.~\ref{fig:ngc1365_halpha}, but for NGC~2992N.}
\label{fig:ngc2992_halpha}
\end{figure}

\section{Analysis and results: NGC~2992}
\label{sec:results2}

\subsection{Direct spectrum}

The direct spectrum of NGC~2992N (Fig.~\ref{fig:spola_ngc2992}) also shows a red continuum with Balmer and forbidden emission lines, the latter lines being more prominent compared to  NGC~1365N.  The broad and narrow lines were separated following the procedure used for NGC~1365N, except that the broad Balmer lines were better fitted with Gaussian profiles. The results are shown in Figs.~\ref{fig:ngc2992_hbeta} and~\ref{fig:ngc2992_halpha}. The narrow lines are unresolved (FWHM $\simeq$ 700 km~s$^{-1}$). The FWHM of the broad Balmer lines is around 2300 km~s$^{-1}$, in agreement with \citet{2021Guolo}. The broad H$\alpha$ and H$\beta$ lines are blueshifted with respect to the narrow lines by $\sim$ 250 and 400 km~s$^{-1}$, respectively. On April 20, 2018, NGC~2992N is in a type~1.8 state given the broad H$\beta$ / [\ion{O}{iii}] ratio \citep{1992Whittle}. From spectra obtained at several epochs between 1978 and 2021, \citet{2021Guolo} showed that NGC~2992N oscillates between types 1.8 and 2, losing and regaining the broad H$\alpha$ emission line. They reported that NGC~2992N was in a type~1.8 state in 2018, in agreement with our results.

The observed H$\alpha$/H$\beta$ broad emission line ratio is equal to 32, which is very high. Adopting a theoretical ratio of 3.06 \citep{2008Dong} and following \citet{2013Momcheva}, we found $E(B-V)$ = 2.0 and $A_V$ = 8 mag, indicating a large extinction. 

\subsection{Polarization}

As in the case of NGC~1365N, the polarization of NGC~2992N (Fig.~\ref{fig:spola_ngc2992}) slowly decreases toward redder wavelengths. At the blue end of our spectra we measure $p \simeq 4.5\%$, while at the red end $p \simeq 3 \%$ (Table~\ref{tab:pola}). However, in contrast with NGC~1365N, the polarization angle is remarkably constant over the full wavelength range : $\theta = 38 \pm 1\degr$. In Table~\ref{tab:pola} we list the polarization measurements obtained between 1980 and 1995, when NGC~2992N was in a type~1.9 or a type~2 state. We also provide our new polarization measurements integrated over similar bandwidths for comparison. At all epochs and for the different AGN states, the polarization angle appears constant within the uncertainties. The polarization degree remains high with the same wavelength dependence, the observed differences being most likely due to a variable contamination by unpolarized light from the host galaxy, which depends on the apertures used for the integration \citep[e.g.,][]{1999Felton}.  Spectropolarimetric observations of NGC~2992N were also obtained in June 2002 by \citet{2004Lumsden}. No broad H$\beta$ was detected in the direct spectrum,  so that NGC~2992N was in a type~1.9 state at that epoch. The polarization degree showed the same wavelength dependence as ours, with an average polarization angle of 35\degr\ \citep{2004Lumsden}. 

The radio emission from  NGC~2992N shows a pair of lobes with an eight-shaped morphology at a position angle of about 160\degr\ \citep{1984Ulvestad,1988Wehrle}. The optical polarization is neither parallel nor perpendicular to that structure, although such alignments are commonly seen in other Seyfert galaxies \citep{1983Antonucci,1984Antonucci}. On the other hand, the polarization is roughly parallel to the dust lane which has a position angle of about 30\degr\ (Fig.~\ref{fig:ngc2992}). On the 1.3~mm maps obtained with the Atacama Large Millimeter Array (ALMA), the AGN appears embedded in the cold dust emission from the highly inclined galaxy disk, which is elongated along a direction with a position angle of about 30\degr\ \citep{2023Zanchettin}. It is therefore most likely that the polarization of NGC~2992N is dominated by dichroic extinction due to aligned dust grains in the host galaxy disk, as already suggested by \citet{1984Antonucci} and \citet{1988Thompson}, and in agreement with the large extinction estimated from the Balmer decrement. This explains why the polarization degree decreases with increasing wavelengths, while the polarization angle remains constant, including in the emission lines. No other polarization mechanism is needed in this case. While dichroic extinction naturally explains the polarization of the narrow lines, it is quite unexpected that the polarization of the emission lines, broad and narrow, is higher than the polarization of the continuum. This can be explained if dilution by unpolarized light is stronger in the continuum than in the emission lines \citep{1988Thompson}. Inhomogeneities in the dust extinction could also be invoked, but the fact that the broad emission lines show high polarizations as the narrow lines which originate from a much larger region, disfavors this interpretation. If this interpretation is correct, one would expect the continuum polarization to be slightly higher in the bright state (type~1.8)  than in the dim states (types~1.9 and~2). Unfortunately, the lower quality of the data previously obtained, as well as the different apertures used to extract the polarization data, precludes any meaningful comparison.  

In summary, dichroic extinction in the NGC~2992 highly inclined disk dominates the polarization of the AGN, overwhelming  polarization due to scattering. The faint broad lines observed in the polarized flux are thus most likely not scattered light, but direct light polarized by the interstellar medium in NGC~2992, the continuum polarization being more diluted than the emission line polarization.

\begin{table}[t]
\caption{Polarization of NGC~2992N}
\label{tab:pola}
\centering
\begin{tabular}{lccccc}
\hline\hline
  Date & Type & $\lambda$ range& $p$  & $\theta$ &  Ref.    \\
(yyymmdd)  &      & (\AA)            & (\%) & (\degr)  &          \\
\hline 
1980-04-08 & 1.9 & 3800-5600  &  3.32$\pm$0.18  & 33$\pm$2   & 1 \\
1980-04-16 & 1.9 & 5300-6600  &  2.84$\pm$0.27  & 31$\pm$3   & 1 \\
1980-04-15 & 1.9 & [\ion{O}{iii}]  &  6.61$\pm$0.59  & 41$\pm$3   & 1 \\
1980-04-15 & 1.9 & H$\alpha$            &  5.68$\pm$0.33  & 41$\pm$1   & 1 \\
1986-02-01 & 2   & B  &  3.34$\pm$0.34  & 39$\pm$3   & 2 \\
1986-02-01 & 2   & V  &  3.16$\pm$0.26  & 36$\pm$3   & 2 \\
1986-02-01 & 2   & R  &  2.40$\pm$0.21  & 38$\pm$3   & 2 \\
1986-02-01 & 2   & J  &  1.27$\pm$0.11  & 36$\pm$3   & 2 \\
1986-02-01 & 2   & H  &  1.18$\pm$0.11  & 38$\pm$3   & 2 \\
1989-03-04 & 2   & 3200-6200  &  3.11$\pm$0.10  & 36$\pm$1   & 3 \\
1990-12-11 & 2   & [\ion{O}{iii}]  &  3.97$\pm$0.41  & 40$\pm$3   & 4 \\
1995-01-30 & 2   & V  &  3.28$\pm$0.39  & 39$\pm$3   & 5 \\
1995-01-30 & 2   & I  &  2.68$\pm$0.25  & 38$\pm$3   & 5 \\
\hline
2018-04-20 & 1.8 & 4620-4820 &  4.56$\pm$0.10  & 38$\pm$1   & 6 \\
2018-04-20 & 1.8 & 5300-6600 &  3.92$\pm$0.03  & 37$\pm$1   & 6 \\
2018-04-20 & 1.8 & 8350-8550 &  2.92$\pm$0.06  & 38$\pm$1   & 6 \\
2018-04-20 & 1.8 & V         &  4.17$\pm$0.03  & 38$\pm$1   & 6 \\
2018-04-20 & 1.8 & R         &  3.65$\pm$0.02  & 37$\pm$1   & 6 \\
2018-04-20 & 1.8 & I         &  2.65$\pm$0.02  & 38$\pm$1   & 6 \\
2018-04-20 & 1.8 & H$\beta$        &  5.27$\pm$0.15  & 38$\pm$1   & 6 \\
2018-04-20 & 1.8 & [\ion{O}{iii}]  &  4.98$\pm$0.15  & 38$\pm$1   & 6 \\
2018-04-20 & 1.8 & H$\alpha$       &  4.81$\pm$0.10  & 38$\pm$1   & 6 \\
2020-01-01 & 1.8 & Gunn $g$  &  3.72$\pm$0.03  & 36$\pm$1   & 7 \\
\hline
\end{tabular}
\tablefoot{The AGN types from 1980 to 1995 and in 2020 were estimated from the compilation of \citet{2021Guolo}. References for the polarization measurements: (1) \citet{1988Thompson}; (2) \citet{1990Brindle}; (3) \citet{1994Kay}; (4) \citet{1992Goodrich}; (5) \citet{1999Felton}; (6) This work - FORS2 ; (7) This work - ISIS. The errors of the polarization angle are rounded to the nearest degree, and to 1\degr\ when smaller than one degree.} 
\end{table}

\section{Conclusions}
\label{sec:conclu}

We  analyzed new (spectro)polarimetric observations of the changing-look AGNs embedded in the inclined spiral galaxies NGC~1365 and NGC~2992. Both AGNs were known as radio emitters, thus making it possible to compare their polarization to their radio axis.

In NGC~1365, the AGN shows polarization properties typical of type~1 Seyferts, i.e., polarization parallel to the radio jet. This intrinsic polarization is modified by the wavelength-dependent dichroic extinction that occurs in the galaxy bar and that rotates the polarization angle at the shortest wavelengths. It would be particularly interesting to catch NGC~1365N in a type~2 state, and see if its polarization shifts to perpendicular or remains parallel to the radio axis. A disappearance of the BLR can also lead to the disappearance of the equatorial scattering region at the origin of the parallel polarization.

NGC~2992, on the other hand, is so inclined that dichroic dust extinction in the disk completely dominates the polarization of the AGN, overwhelming  any polarization due to scattering. The polarization properties thus remain essentially constant between the different AGN states. The faint broad lines observed in the polarized flux are thus most likely not scattered light. The differential dilution of the continuum and the narrow-line polarizations may explain the unusually high polarization of the lines. This interpretation could be further explored by measuring subtle polarization changes between the different changing-look states.

\begin{acknowledgements}
This research has made use of the NASA/IPAC Extragalactic Database, which is funded by the National Aeronautics and Space Administration and operated by the California Institute of Technology. D.H. is Research Director at the F.R.S.-FNRS, Belgium. BAG was funded by the European Union ERC-2022-STG - BOOTES - 101076343. Views and opinions expressed are however those of the author(s) only and do not necessarily reflect those of the European Union or the European Research Council Executive Agency. Neither the European Union nor the granting authority can be held responsible for them. J.A.P. acknowledges financial support from the Spanish Ministry of Science and Innovation (MICINN) through the Spanish State Research Agency, under Severo Ochoa Centres of Excellence Programme 2020-2024 (CEX2019-000920-S).
\end{acknowledgements}

\bibliographystyle{aa}
\bibliography{references}

\begin{appendix}
  
\section{NGC~1365N in 2024} 
\label{appa}

\begin{figure}[h]
\centering
\resizebox{0.9\hsize}{!}{\includegraphics*{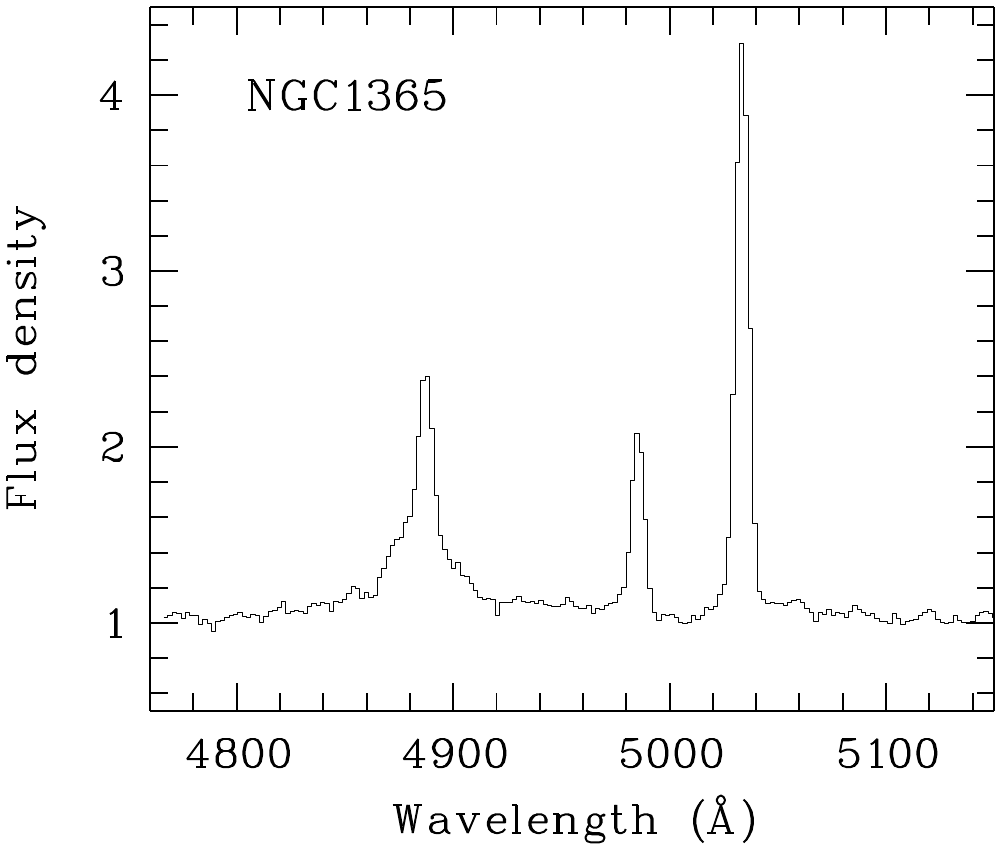}}
\caption{The H$\beta$ + [\ion{O}{iii}] spectral region of NGC~1365N on February 21, 2024. The flux density is given in arbitrary units.}
\label{fig:efosc}
\end{figure}

A series of four spectra of the nucleus of NGG~1365 were obtained on February 21, 2024, with the ESO new technology telescope (NTT) equipped with the ESO faint object camera and spectrograph (EFOSC). The data were retrieved from the ESO archive (program ID: 60.A-9501; PI: La Silla Observing School). The observations were carried out with grism~\#18, that covers the wavelength range 4120-7420 \AA. The spectra were bias-subtracted, flat-fielded, and wavelength-calibrated using standard procedures from the ESO-MIDAS reduction system. The four spectra were finally co-added. The spectral region including the H$\beta$ and [\ion{O}{iii}] emission lines is shown in Fig.~\ref{fig:efosc}, clearly showing the H$\beta$ broad component.

\section{NGC1365N polarization in the $q$, $u$ plane} 
\label{appb}

\begin{figure}[t]
\centering
\resizebox{0.9\hsize}{!}{\includegraphics*{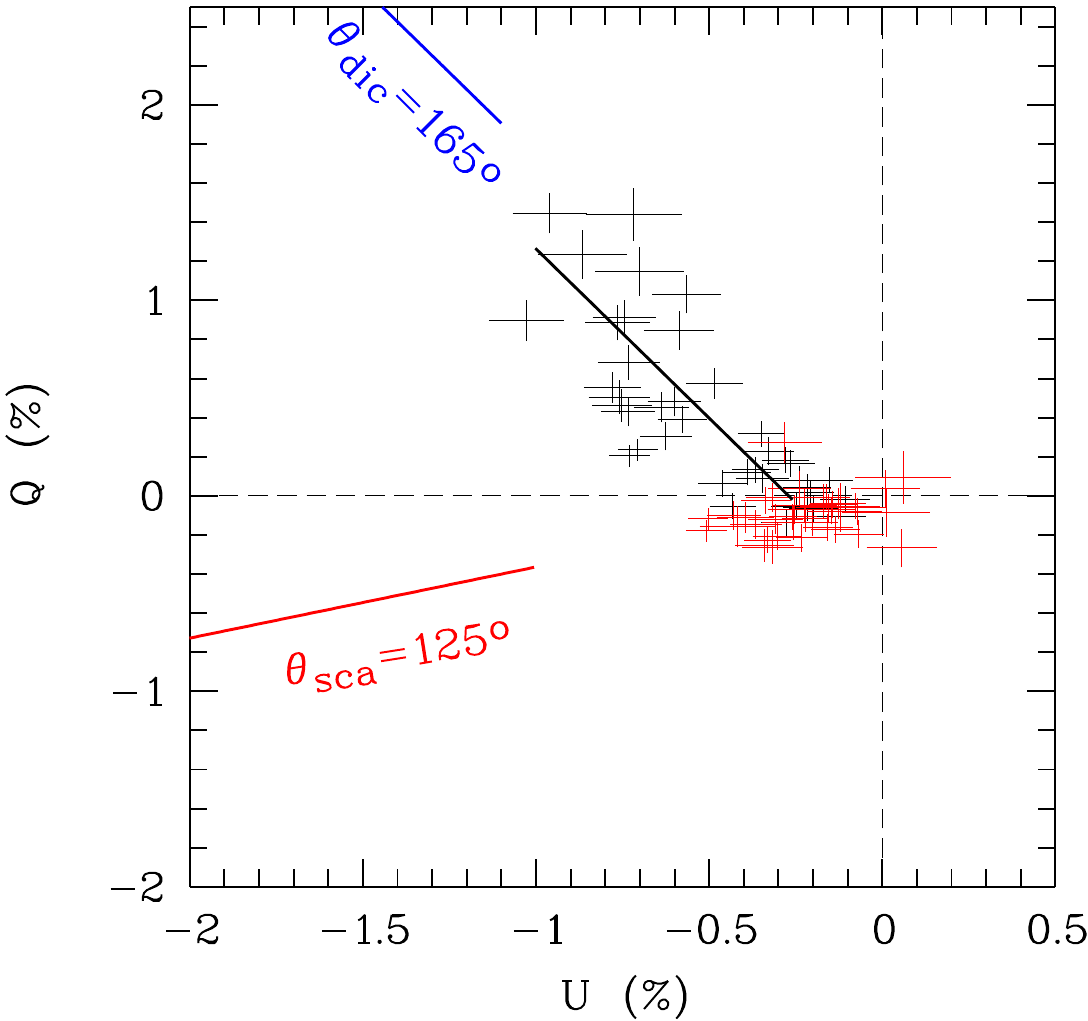}}
\caption{Polarization spectrum of NGC 1365N, binned over 30 spectral pixels (100 \AA), plotted in the normalized Stokes $q$, $u$ plane. The black crosses represent the data, with their error bars. The solid black line represents the polarization model fitted to the data. The model is a combination of polarization scattering at $\theta \simeq 125\degr$ and polarization due to dichroic extinction at $\theta \simeq 165\degr$, the individual components being indicated by the red and blue solid lines, respectively. The red crosses represent the measured $q$, $u$ parameters corrected for dichroic extinction, with error bars.}
\label{fig:ngc1365_uq}
\end{figure}

We modeled the polarization observed in NGC~1365N with a contribution due to scattering assumed to vary linearly between $p_{\rm sca}^{\rm blue}$ and $p_{\rm sca}^{\rm red}$ over the observed wavelength range, and a contribution due to dichroic extinction assumed to follow a Serkowski-type law: $p_{\rm dic} = p_{\rm max} \exp [-K \ln ^{2} (\lambda_{\rm max}/\lambda)]$  \citep{1975Serkowski}. The modeled $q_{\rm mod}$ and $u_{\rm mod}$ are computed as follows:
\begin{eqnarray}
 q_{\rm mod} & = & p_{\text{sca}} \; \cos(2 \, \theta_{\text{sca}}) + p_{\text{dic}} \; \cos (2 \, \theta_{\text{dic}})\\
 u_{\rm mod} & = & p_{\text{sca}} \; \sin(2 \, \theta_{\text{sca}}) + p_{\text{dic}} \; \sin (2 \, \theta_{\text{dic}}) \; ,
\end{eqnarray}
and obtained by fitting the measured polarization in the $q$, $u$ plane (Fig.~\ref{fig:ngc1365_uq}). $p_{\rm sca}^{\rm blue}$, $p_{\rm sca}^{\rm red}$,  $p_{\rm max}$, $K$, and $\lambda_{\rm max}$ are free parameters. We imposed $p_{\rm sca}^{\rm blue} \gtrsim p_{\rm sca}^{\rm red}$, and constrained $K$ and $\lambda_{\rm max}$ in the ranges [0, 5] and [3000\AA , 9000\AA], respectively. $\theta_{\text{dic}}$ = 165\degr\ and $\theta_{\text{sca}}$ = 125\degr\ are fixed. The best fit, shown in Fig.~\ref{fig:ngc1365_uq}, is obtained with $p_{\rm sca}^{\rm blue} \simeq p_{\rm sca}^{\rm red} \simeq 0.24\%$,  $p_{\rm max} = 1.83\%$, $K = 5$, and $\lambda_{\rm max} = 3840\AA $. The observed $q_{\rm obs}$ and $u_{\rm obs}$ values were then corrected for the dichroic extinction, so as to isolate the scattering component. This component roughly follows a straight line that crosses the origin in the $q$, $u$ plane (red crosses in Fig.~\ref{fig:ngc1365_uq}).

\end{appendix}

\end{document}